\documentstyle[12pt]{article} 

\begin{document}

\begin{flushright}
hep-ph/9906212
\end{flushright}
\vspace{1cm}
\begin{center}

{\large{\bf Some Theoretical Results on $V_{ub}$ and $V_{cb}$}}
\footnote{Contribution to the Workshop on the Derivation of $|V_{cb}|$ and
$|V_{ub}|$: Experimental Status and Theory Uncertainties, CERN, 
May 28 - June 2, 1999} \\
\vspace{0.7cm}
Changhao Jin\\
{\cal School of Physics, University of Melbourne\\
Parkville, Victoria 3052, Australia}\\
Email: jin@physics.unimelb.edu.au
\vspace{0.4cm}

\end{center}

I will highlight some of our theoretical results on $V_{ub}$ and $V_{cb}$
from inclusive decays of $b$ hadrons. Please refer to the original papers for
details.  
\section{$V_{\lowercase{ub}}$ from inclusive charmless semileptonic decays of 
$\lowercase{b}$ hadrons}

\subsection{A new method --- a model-independent determination}

It was proposed \cite{new} recently to use the decay distribution in terms of 
the observable $\xi_u=(q^0+|{\bf q}|)/M_B$ in the $B$ rest frame to measure 
$V_{ub}$, where $q$ is the momentum transfer to the lepton pair.

This decay spectrum is unique in that the tree-level and virtual gluon 
processes $b\to u\ell\nu$ at the parton level generate a trivial $\xi_u$ 
spectrum --- a discrete line at $\xi_u=m_b/M_B$, solely on kinematic grounds. 
Two distinct effects, gluon bremsstrahlung and hadronic bound state effects,
spread out the spectrum, but most of the decay rate remains at large $\xi_u$.
Consequently, about $99\%$ of the $b\to u$ events pass the kinematic cut
$\xi_u > 1-M_D/M_B$, where no $b\to c$ transition is allowed. This 
discrimination between $b\to u$ signal and $b\to c$ background is even more 
efficient than the cut on the hadronic invariant mass.

Because of heaviness of the decaying hadron, the light-cone expansion is
applicable to inclusive $B$ decays \cite{jp,vcb,lepton,baryon,rare}. 
The leading nonperturbative QCD effect is attributed to the distribution 
function, defined as Fourier transformation of the matrix element of the
non-local $b$ quark operators separated along the light cone.
The spectrum $d\Gamma/d\xi_u$ is directly proportional to the 
distribution function \cite{new}. The
detailed form of the distribution function is not known. However, the 
normalization of it is exactly known due to $b$ quantum number conservation.
Using the known normalization, the dependence on the distribution function can
be eliminated in the weighted integral of the decay spectrum:
\begin{equation}
\int_0^1 d\xi_u \frac{1}{\xi_u^5}\frac{d\Gamma}{d\xi_u}= \frac{G_F^2M_B^5}
{192\pi^3}|V_{ub}|^2 .
\end{equation}
Thus a measurement of the above weighted integral of the $\xi_u$ spectrum 
determines $V_{ub}$. This method is based on the light-cone expansion, which
is, in principle, model independent as it is in deep inelastic scattering.   
Note that this method does not rely on the heavy quark effective theory.

Therefore, at least potentially, this theoretically sound, clean and 
experimentally efficient method allows for a model-independent determination 
of $V_{ub}$ with a minimum overall (experimental and theoretical) error.
By this method the dominant hadronic uncertainty associated with the 
distribution function is avoided.
The residual hadronic uncertainty due to higher-order, power-suppressed
corrections of order $O(\Lambda^2_{\rm QCD}/M_B^2)$ is expected to be at the 
level of $1\%$. The perturbative 
corrections are calculable. The study of these remaining theoretical 
uncertainties is in progress. The precision of this determination of $V_{ub}$
will mainly depend on its experimental feasibility. This method appears 
quite feasible
by the similar techniques used to measure the inclusive charmless semileptonic
branching ratio of $b$ hadrons at LEP. This method may be the best one 
available for the $V_{ub}$ determination.

\subsection{From the inclusive charmless semileptonic branching ratio}

I have calculated \cite{vub} the inclusive charmless semileptonic $B$ decay 
width in the approach \cite{jp,vcb,lepton,baryon,rare}
based on light-cone expansion and heavy quark effective theory.
This approach is from first principles and the nonperturbative QCD effect can 
be computed in a systematic way. Additional properties of the distribution 
function were deduced from the heavy quark effective theory, which impose
strong constraints on the functional form of it. 
This allows for a largely model-independent determination of $V_{ub}$ from
the inclusive charmless semileptonic branching ratio of $b$ hadrons measured
at LEP.

A crucial observation is that both dynamic and kinematic effects of 
nonperturbative QCD must be taken into account. The latter results in the 
extension of phase space from the quark level to the hadron level, which 
obviously increases the decay width. It turns out that the net effect of
nonperturbative QCD enhances the semileptonic decay width.

The heavy quark expansion approach \cite{hqe} fails
to take into account the kinematic effect of nonperturbative QCD and, as a
result, the calculation \cite{ural} of the decay width in this approach lead 
to a higher value of $V_{ub}$. This failure is a consequence of the 
theoretical limitations in the heavy 
quark expansion approach: the operator product expansion breaks down for 
low-mass final hadronic states; the truncation of the expansion enforces the
use of quark kinematics rather than physical hadron kinematics. These 
theoretical
limitations were already indicated by the $\tau(\Lambda_b)/\tau(B_d)$ 
measurements. The recent CLEO analysis \cite{cleo}
of the hadronic mass and lepton energy 
moments in $B\to X_c\ell\nu$ may also hint at such limitations if the 
experiment is correct. The preliminary experimental result shows \cite{cleo} 
an inconsistency between the values of the heavy quark expansion parameters
extracted, respectively, from the measured hadronic mass moments and from
the measured lepton energy moments.

Moreover, the interplay between nonperturbative and perturbative QCD effects 
has been accounted for in our approach, since confinement implies that free 
quarks are not asymptotic states of the theory.

\subsection{From the hadronic invariant mass spectrum}

I have analysed \cite{mx} the hadronic invariant mass spectrum in the QCD-based
approach \cite{jp,vcb,lepton,baryon,rare}. I found that the theoretical error 
on $V_{ub}$ depends strongly on the hadronic 
invariant mass cutoff. The higher it can be experimentally made to be, the 
smaller the theoretical error on $V_{ub}$.

The hadronic invariant mass spectrum has also been analysed \cite{falk}
in the approach \cite{resum} based on the resummation of the heavy quark 
expansion. The distinct approximations are made in this approach. A comparison
found \cite{vub,rare} that this approach contains less information on 
nonperturbative QCD in the leading approximation than the 
approach \cite{jp,vcb,lepton,baryon,rare} based on light-cone expansion.  

\subsection{From the lepton energy endpoint spectrum}

This determination of $V_{ub}$ has statistical power. Our analysis in the 
QCD-based approach showed \cite{lepton} that the theoretical uncertainty on 
$V_{ub}$ from the lepton energy endpoint spectrum is under control. 

A key step towards the improvement of the theoretical uncertainties on $V_{ub}$
from the
inclusive charmless semileptonic branching ratio, the hadronic invariant mass
spectrum or the lepton energy endpoint spectrum is a direct extraction of the 
distribution function from experiment. It was pointed out \cite{new,rare} that
the nonperturbative distribution function can be directly extracted by 
measuring either the spectra in $\xi_f=(q^0+\sqrt{|{\bf q}|^2+m_f^2})/M_B$ 
$(f= u, c)$ in inclusive semileptonic $B$ decays or the photon
energy spectrum in inclusive radiative $B$ decays. 

\section{Model-independent determinations of the ratios of the CKM matrix 
elements}

It was found \cite{rare} that the distribution function is universal in the 
sense that the same distribution function encodes the leading nonperturbative 
QCD contributions to inclusive semileptonic $B$ decays as well as inclusive
radiative $B$ decays. 
It was proposed \cite{rare} that a model-independent determination of the 
ratio $|V_{ub}/V_{ts}|$ can be obtained by measuring the ratio of the $\xi_u$
spectrum in $B\to X_u\ell\nu$ and the photon energy spectrum in 
$B\to X_s\gamma$, since the universal distribution function cancels in the
ratio, 
$[d\Gamma(B\to X_u\ell\nu)/d\xi_u]/
[d\Gamma(B\to X_s\gamma)/dE_\gamma]|_{E_\gamma= M_B\xi_u/2}$.
By the similar methods one can also obtain model-independent determinations of
$|V_{ub}/V_{cb}|$ \cite{new} and $|V_{cb}/V_{ts}|$ \cite{rare}.
These methods depend on the validity of universality of the distribution 
function, which can be tested experimentally.

\section{$V_{\lowercase{cb}}$ from inclusive charmed semileptonic decays of 
$\lowercase{b}$ hadrons}

\subsection{From the inclusive semileptonic branching ratio}

I have calculated \cite{vcb} the semileptonic decay width of the $B$ meson,
which can be used to gain a largely model-independent determination of 
$V_{cb}$. It was also shown that it is important to include the kinematic 
nonperturbative QCD effect, as in the $b\to u$ case discussed above. I 
found \cite{vcb}
that the semileptonic decay width is enhanced by long-distance strong 
interactions,
in contrast to the result of the heavy quark expansion where a reduction of
the free quark decay width is claimed. The primary reason for the difference 
is that 
the heavy quark expansion approach has to use the quark-level kinematics 
rather than the hadron-level kinematics, as mentioned above.
Consequently, compared with the light-cone 
approach \cite{jp,vcb,lepton,baryon,rare},
the inclusive rate calculated in the heavy quark expansion 
approach leads to a larger gap between the inclusive and exclusive 
determinations of $|V_{cb}|$.
 
Our prediction for the lepton energy spectrum was found \cite{lepton} to be in 
good agreement with the experimental data. This experimental test increases 
our confidence in the determination of $V_{cb}$.

\subsection{A new method}

It was proposed \cite{new} to use the $\xi_c$ spectrum to obtain a 
model-independent determination of $V_{cb}$. The idea is the same as the use
of the $\xi_u$ spectrum to determine $V_{ub}$ discussed in Section 1.1.
However, this way of determining $V_{cb}$ may still suffer from large 
theoretical systematic error. For $B\to X_c\ell\nu$, the maximum momentum 
transfer squared is $q_{\rm max}^2= (M_B-M_D)^2$. 
This means that $q^2$ is not large enough to 
neglect the higher order corrections. Actually the semileptonic $b\to c$ decay
rate is dominated by a few exclusive decay modes ($D, D^\ast$ and $D^{(\ast)}
\pi$), which suggests that the light-cone picture cannot be valid point by 
point. The theoretical prediction in the light-cone expansion refers only to
the smeared spectrum. A related problem is the uncertainty in the charm quark
mass.

The method proposed for a model-independent determination of $V_{ub}$ is, on
the other hand, theoretically very reliable. The light-cone expansion works
much better for $B\to X_u\ell\nu$ because a much larger momentum transfer with
the maximum $q_{\rm max}^2= M_B^2$ can occur in $B\to X_u\ell\nu$ than in 
$B\to X_c\ell\nu$. Many final hadronic states contribute to the $\xi_u$ 
spectrum above the charm threshold, without any preferential weighting towards
the low-lying resonance states. Both theoretical and experimental situations
in this way of determining $V_{ub}$ are so attractive that the feasibility of 
the experiment is worth investigating.

\vspace{0.4cm}
{\it Note added.} After this note was completed, a paper by 
Uraltsev \cite{comm} appeared, in which the work of Refs.~\cite{vcb,vub} was 
criticized \footnote{The same criticism appeared in \cite{comm1}, which is
addressed here. It is worth reminding that the semileptonic decay width
Eq.~(1) in \cite{comm} (or Eq.~(13) in \cite{comm1}) was derived using quark
kinematics from the heavy quark expansion. Contrary to claims in 
Ref.~\cite{comm1}, there exists no rigorous proof in the literature (including
Refs.~[3,4,12,13] given in \cite{comm1}) which demonstrates that Eq.~(1) in
\cite{comm} (or Eq.~(13) in \cite{comm1}) recovers the full decay width from
hadron kinematics. The sum rules are a key ingredient of the proof claimed 
in Ref.~\cite{comm1}. These sum rules were obtained by assuming that the 
moments
of the structure functions are identical to the moments of the structure 
functions at the quark level (see, e.g., Eq.~(102) of Ref.~[4] 
in \cite{comm1}).
However, this assumption is not valid and, as a matter of fact, the sum rules
were obtained still using quark kinematics under the assumption. The claimed
proof is therefore incorrect. Rather, it is logically apparent that the
kinematic nonperturbative effect due to the extention from quark phase space
to hadron phase space cannot be included in the decay width calculations in
quark phase space itself that lead to Eq.~(1) in \cite{comm} and Eq.~(13) 
in \cite{comm1}. Quantitatively the significant difference shown 
in \cite{vcb,vub} between the semileptonic widths calculated in the light-cone
approach using hadron kinematics and in the heavy quark expansion approach
using quark kinematics confirms that the kinematic nonperturbative QCD 
contributions are missed in the heavy quark expansion approach.}. 
It has been observed in Refs.~\cite{vcb,vub} that the
kinematic nonperturbative QCD effects are missed in the heavy quark expansion 
approach and must be incorporated additionally. The author of Ref.~\cite{comm} 
disproves that observation by criticizing the approach in 
Refs.~\cite{vcb,vub}. However, 
it should be stressed that that observation does not depend on any specific 
approach. The total decay rate receives an enhancement when the phase space
is extended from the quark level determined by the $b$ quark mass to the 
hadron level determined by the $B$ meson mass. This is physically quite 
obvious and general, without the intervention of any specific theoretical
approach. The heavy quark expansion approach has to use the quark-level phase 
space, while the $B$-meson decay rate should be calculated using the physical
hadron-level phase space.
As a consequence, there is rate missing in the calculation of
the inclusive charmless semileptonic decay rate in \cite{ural} in the 
heavy quark expansion approach.

Let me next turn to the issue of the theoretical foundation of the approach
in \cite{vcb,vub}. The light-cone expansion has long been recognized as the 
theoretical
foundation for the description of deep inelastic scattering processes that are
dominated by light-cone singularities. The same formalism is at the basis of
the approach \cite{jp,vcb,lepton,baryon,rare} to inclusive $B$ decays.
Inclusive semileptonic $B$ decays 
involve large momentum transfer in most of phase space. The light-cone
expansion is applicable to the inclusive decays. On the other hand, the heavy
quark expansion for inclusive semileptonic $B$ decays is grounded in the 
operator
product expansion in the case of large energy release. Making a serious 
scrutiny into their formulations, one can realize that as concerns the 
theoretical
foundation the approach based on the light-cone expansion in \cite{vcb,vub}
is not less firm at all than the heavy quark expansion approach.

Actually these two approach tackle nonperturbative QCD effects in the 
different ways. The discrepancy between their results is an inevitable 
consequence of the difference between the underlying methods. 
Given the theoretical
problems of the heavy quark expansion approach mentioned previously, it is
not justified to regard such a discrepancy as the deficiency of the approach 
based on the light-cone expansion. The discrepancy could be just a reflection 
of the merit of the light-cone approach. As an example, Eq.~(14) in 
Ref.~\cite{comm}
results from the first three terms in the moment expansion \cite{jp}
of the distribution function in the light-cone approach:
\begin{equation}
f(\xi)=\sum_{n=0}^{\infty}\frac{(-1)^n}{n!}M_n(\frac{m_b}{M_B})
\delta^{(n)}(\xi-\frac{m_b}{M_B}) .
\end{equation}
In fact,
the truncation of this expansion is illegal in the light-cone approach.
Thus Eq.~(14) given in \cite{comm} is not a true result of the light-cone 
approach. Nevertheless, assuming that they are comparable the emerging 
discrepancy between 
Eq.~(14) and the heavy quark expansion result Eq.~(1) shown in \cite{comm} 
is conceivable
and not surprising, and not in a problem with the light-cone approach itself.

In the free quark limit, the $B$ meson and the $b$ quark in it move together
with the same velocity. Eq.~(2) of Ref.~\cite{vub} is the expression for the
inclusive charmless semileptonic decay width of the $B$ meson at rest, 
 from the light-cone expansion.
In the free quark limit, corresponding to the limit 
$f(\xi)\to\delta(\xi-m_b/M_B)$, it correctly reproduces the decay width of
the free $b$ quark at rest. The claimed meaning of the heavy quark expansion
correction in Eq.~(1) of Ref.~\cite{comm} corresponding to a free quark moving
with the small velocity is self-contradictory and not a result of QCD.

A related point is that if true the results from QCD in $1+1$ dimensions may
provide some insights when we have to deal with real systems that are not
simple, but cannot be regarded as the truth or proof in real QCD.

I would conclude that it is premature to dismiss the results of 
Refs.~\cite{vcb,vub} on the basis of the work in \cite{comm}.
It is an indisputable fact that the kinematic nonperturbative QCD effects
identified in \cite{vcb,vub} are missed in the heavy quark expansion approach.

\vspace{0.4cm}
{\bf Acknowledgements:} I am indebted to Emmanuel Paschos for collaboration.
I would like to thank the organizers for inviting me to have a contribution. 
This work is supported by the Australian Research Council.


\begin{thebibliography}{23}
\bibitem{new} C.H. Jin, Mod. Phys. Lett. A 14 (1999) 1163; hep-ph/9808313.
\bibitem{jp} C.H. Jin and E.A. Paschos, in {\it Proceedings of the 
International 
Symposium on Heavy Flavor and Electroweak Theory}, Beijing, China, 
1995, edited by C.H. Chang and C.S. Huang (World Scientific, Singapore,
1996), p.~132; hep-ph/9504375.
\bibitem{vcb} C.H. Jin, Phys. Rev. D 56 (1997) 2928.
\bibitem{lepton} C.H. Jin and E.A. Paschos, Eur. Phys. J. C 1 (1998) 523.
\bibitem{baryon} C.H. Jin, Phys. Rev. D 56 (1997) 7267.
\bibitem{rare} C.H. Jin, hep-ph/9903510.
\bibitem{vub} C.H. Jin, Phys. Lett. B 448 (1999) 119.

\bibitem {hqe}
J. Chay, H. Georgi and B. Grinstein, Phys. Lett. B 247 (1990) 399;\\
I.I. Bigi, N.G. Uraltsev and A.I. Vainshtein, Phys. Lett. B 293 (1992) 430;
B 297 (1993) 477(E);\\
I.I. Bigi, M.A. Shifman, N.G. Uraltsev and A.I. Vainshtein, Phys.
Rev. Lett. 71 (1993) 496;\\
A.V. Manohar and M.B. Wise, Phys. Rev. D 49 (1994) 1310;\\
B. Blok, L. Koyrakh, M.A. Shifman and A.I. Vainshtein, Phys. Rev.
D 49 (1994) 3356; D 50 (1994) 3572(E).

\bibitem{ural} N. Uraltsev, Int. J. Mod. Phys. A 11 (1996) 515;\\
I. Bigi, M. Shifman, N. Uraltsev, Annu. Rev. Nucl. Part. Sci. 47 (1997) 591.

\bibitem{cleo} CLEO Collaboration, J. Bartelt {\it et al.}, CLEO CONF 98-21.

\bibitem{mx} C.H. Jin, Phys. Rev. D 57 (1998) 6851.

\bibitem{falk} A.F. Falk, Z. Ligeti, M.B. Wise, Phys. Lett. B 406 (1997) 225;\\
R.D. Dikeman, N. Uraltsev, Nucl. Phys. B 509 (1998) 378;\\
I. Bigi, R.D. Dikeman, N. Uraltsev, Eur. Phys. J. C 4 (1998) 453.

\bibitem{resum} M. Neubert, Phys. Rev. D 49 (1994) 3392; 4623;\\
T. Mannel and M. Neubert, Phys. Rev. D 50 (1994) 2037;\\
I.I. Bigi, M.A. Shifman, N.G. Uraltsev and A.I. Vainshtein,
Int. J. Mod. Phys. A 9 (1994) 2467.

\bibitem{comm} N. Uraltsev, hep-ph/9905520.
\bibitem{comm1} I.I. Bigi, hep-ph/9907270.
\end{thebibliography}
\end{document}